\documentclass[
reprint,
superscriptaddress,
amsmath,amssymb,
aps,
prb,
]{revtex4-2}
\usepackage{physics}
\usepackage{graphicx}
\usepackage{dcolumn}
\usepackage{bm}

\begin{document}

\title[]{\large Microscopic Theory of Polariton Group Velocity Renormalization}

\author{Wenxiang Ying}
\email{W. Y. and B. X. K. C. contributed equally to this work.}
\affiliation{Department of Chemistry, University of Rochester, 120 Trustee Road, Rochester, NY 14627, USA}

\author{Benjamin X. K. Chng}
\email{W. Y. and B. X. K. C. contributed equally to this work.}
\affiliation{Department of Physics and Astronomy, University of Rochester, Rochester, NY 14627, USA}

\author{Milan Delor}
\affiliation{Department of Chemistry, Columbia University, New York, NY 10027, USA}

\author{Pengfei Huo}
\email{pengfei.huo@rochester.edu}
\affiliation{Department of Chemistry, University of Rochester, 120 Trustee Road, Rochester, NY 14627, USA}
\affiliation{The Institute of Optics, Hajim School of Engineering, University of Rochester, Rochester, NY 14627, USA}
\affiliation{Center for Coherence and Quantum Optics, University of Rochester, Rochester, New York 14627, USA}


\begin{abstract}
Cavity exciton-polaritons exhibit ballistic transport and can achieve a distance of 100 $\mu $m in one picosecond. This ballistic transport significantly enhances mobility compared to that of bare excitons, which often move diffusively and become the bottleneck for energy conversion and transfer devices. Despite being robustly reproduced in experiments and simulations, there is no comprehensive microscopic theory addressing the group velocity of polariton transport, and its renormalization due to phonon scattering while still preserving this ballistic behavior. In this work, we develop a microscopic theory to describe the group velocity renormalization using a finite-temperature Green's function approach. Utilizing the generalized Holstein-Tavis-Cummings Hamiltonian, we analytically derive an expression for the group velocity renormalization and find that it is caused by phonon-mediated transitions from the lower polariton (LP) states to the dark states, then scattering from dark states back to LP. The dark states do not have to be populated in this process, serving as the virtual state for super-exchange (especially true for a large light-matter detuning). The theory predicts that the magnitude of group velocity renormalization scales linearly with the phonon bath reorganization energy under weak coupling conditions (perturbative regime for exciton-phonon coupling) and also linearly depends on the temperature in the high-temperature regime. These predictions are numerically verified using quantum dynamics simulations, demonstrating quantitative agreement. Our findings provide theoretical insights and a predictive analytical framework that advance the understanding and design of cavity-modified semiconductors and molecular ensembles, opening new avenues for engineered polaritonic devices.
\end{abstract}

\maketitle

Recent experiments~\cite{Forrest_AM2020, berghuis2022controlling, Musser_2022, Milan_NC2023, Milan_NL2023, Tal_NM2023, Tal_NM2024} have shown that exciton transport in semiconductors can be significantly enhanced by coupling these excitons to confined electromagnetic modes inside an optical cavity. By forming cavity exciton-polaritons, the electronic excitation is capable of traversing long distances ballistically at a high group velocity $v_{g}$. This novel strategy of cavity-enhanced ballistic exciton energy transport allows devices to bypass the intrinsic bottleneck of diffusive transport, offering a paradigm shift in fundamental energy science and device applications such as halide perovskite~\cite{xu2023ultrafast} and light-emitting diode displays~\cite{gather2011white, reineke2009white, reineke2013white}. The high group velocity mainly arises from the large curvature of the dispersion curve of the polariton bands (compared to the pure-matter band). Due to polariton-phonon interactions, experiments~\cite{Milan_NC2023} have shown a further reduction in the group velocity (often referred to as group velocity renormalization) when increasing the excitonic fraction of the polariton, thus deviating from the derivative of the polariton dispersion curve which equates to the upper limit of the group velocity. 

Recent progress in the theoretical understanding of polariton transport~\cite{Pupillo_2015, Zhou_PRA2023, Ribeiro_JCP2023, Ribeiro_Nanophotonics24, Cao_PRL2023, Cao_arXiv2024, berghuis2022controlling, sokolovskii2023multi, chng2024transport} have emerged through numerical simulations~\cite{Tal_NM2023, sokolovskii2023multi,tichauer2023tuning,chng2024transport} and theoretical models~\cite{Tal_NM2023,Ribeiro_JCP2023,Zhou_PRA2023}, providing valuable insights into this complex phenomenon.
In the framework of theoretical models, two prevailing hypotheses for $v_{g}$ renormalization have been presented. 
One is the thermally activated scattering theory in Ref.~\citenum{Tal_NM2023}, which posits that there will be a quasi-equilibrium between the polariton band and the dark exciton states.
Under this theory, ballistic transport occurs only during the period when the system is in the polariton band (see the detail of the theory in the Supplementary Information of Ref.~\citenum{Tal_NM2023}). As such, $v_{g}$ is reduced, and the extent of the renormalization depends on the energy difference between the polariton band and the dark excitons. A similar hypothesis is also proposed in Ref.~\citenum{Musser_2022}. The second hypothesis is the transient localization hypothesis proposed in Ref.~\cite{Milan_NC2023}, which arises from the interpretation of trajectory results in the Ehrenfest mixed-quantum-classical (MQC) simulations. According to this hypothesis, the polariton wavepacket predominantly exhibits ballistic coherent transport, but the wavepacket becomes transiently localized due to phonon coupling. This hypothesis explains the group velocity renormalization and the ballistic transport concurrently and can be examined from the trajectories obtained from the mixed quantum-classical simulations directly. Despite these promising developments, there is no microscopic theory, to the best of our knowledge, that quantitatively describes $v_{g}$-renormalization and shows how $v_{g}$-renormalization depends on exciton-phonon coupling (reorganization energy $\lambda$), temperature $T$, exciton fraction in the polariton, {\it etc}. 


In this work, we develop a microscopic theory using a field-theoretic approach to explain the polariton $v_{g}$-renormalization due to polariton-phonon interactions. By utilizing the polariton Green's functions, we derive the modified band structure for polaritons, which results in a renormalized polariton group velocity. Our theory indicates that within the lower polariton branch, the system manifests a phonon-mediated attractive interaction between the polaritons, thus slowing down the band-like transport. The theory predicts that the extent of modification scales linearly with the phonon bath reorganization energy $\lambda$, and similarly, displays a linear temperature dependence in the high-temperature regime. We also show that the theoretical predictions are in quantitative agreement with numerical results based on MQC simulations~\cite{chng2024transport}.

\section*{Results and Discussions}
{\bf Model System}. We use the Generalized Holstein-Tavis-Cummings (GHTC) Hamiltonian~\cite{herrera2018theory, qiu2021molecular, Arkajit_Chemrev_2023,taylor2024light} to describe $N$ excitons interacting with $\mathcal{M}$ cavity modes, and $N \gg {\mathcal M}$ in line with typical experimental conditions~\footnote{Typically, one estimates $N/\mathcal{M}\sim 10^6-10^9$ for systems used in experiments~\cite{Feist_NJP2015}}. The total Hamiltonian can be written in the form of the system-bath model and is expressed as $\hat{H} = \hat{H}_\text{S} + \hat{h}_\text{B} + \hat{H}_\text{SB}$. The system Hamiltonian $\hat{H}_\mathrm{S}$ consists of the excitonic degrees of freedom (DOF) and the photonic DOF of the cavity. Each exciton is modeled as an effective two-level system that consists of the ground state $|g_{n}\rangle$ and excited state $|e_{n}\rangle$ (for the $n_\mathrm{th}$ exciton). Without making the long-wavelength approximation~\cite{taylor2024light}, $\hat{H}_\text{S}$ is expressed as follows,
\begin{align} \label{eq:Hs_site}
    \hat{H}_\text{S} = &\hbar \omega_0 \sum_{n=1}^N \hat{\sigma}_n^{\dagger} \hat{\sigma}_n +  \sum_{k}^{\mathcal M} \hbar\omega_{k} \hat{a}^{\dagger}_k \hat{a}_k \\
    & +  \sum_{k}  \sum_{n=1}^N \hbar g_{k}\left[\hat{a}^{\dagger}_k \hat{\sigma}_n e^{- i {k}_{\parallel}\cdot {x}_{n}} + \hat{\sigma}_n^{\dagger} \hat{a}_k e^{i {k}_{\parallel}\cdot {x}_{n}} \right], \notag
\end{align}
where $\hat{\sigma}_n^{\dagger}=|e_{n}\rangle\langle g_{n}|$ and $\hat{\sigma}_n=|g_{n}\rangle\langle e_{n}|$ are the creation and annihilation operators of the $n_\mathrm{th}$ molecule's exciton, and $\omega_0$ is the excitation energy between the molecule's ground and excited state. Further, $\hat{a}_{k}$ and $\hat{a}_{k}^{\dagger}$ are the photonic field annihilation and creation operators for mode $k$ whose frequency is $\omega_k$.

For Fabry-P\'erot (FP) cavities, the dispersion is
\begin{align} \label{eq:wk-cavity}
    \omega_{k}(k_{\parallel}) = c\sqrt{k^2_{\perp} + k^2_{\parallel}},
\end{align}
where $c$ is the speed of light in vacuum.
When $k_{\parallel}=0$, the photon frequency is $\omega_\mathrm{c}\equiv\omega_{k}(k_{\parallel}=0)=ck_{\perp}$.  
The second line of Eq.~\ref{eq:Hs_site} represents light-matter interaction, where $g_{k} = g_\text{c} \sqrt{(\omega_{k} / \omega_\text{c})} \cos\theta$ is the $k$-dependent light-matter coupling strength~\cite{Milan_NC2023}, and $\tan \theta =k_{\parallel} / k_{\perp}$ is the incident angle. Further, $x_n$ is the position of the $n_\mathrm{th}$ exciton. We consider the cavity modes inside the same simulation box as the excitons, with total size $NL$ along the $k_{\parallel}$ direction ($L=x_{n}-x_{n-1}$). 

\begin{figure}[htbp]
    \centering
    \includegraphics[width=1.0\linewidth]{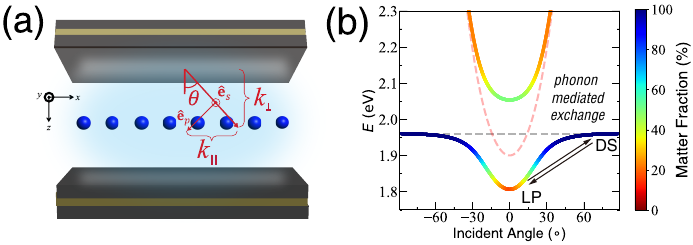}
    \caption{{\bf Schematics of the GHTC model and band structure}. ({\bf a}) Schematics of the model setup. Inside an optical cavity, the separated molecules collectively interact with many cavity modes. ({\bf b}) Polariton band structure, where the matter fraction is shown in terms of colorbar. The dashed lines are the bare photon (red) and matter (silver) dispersions, respectively. The phonon-mediated exchange effect between the lower polariton (LP) and the dark states (DS) manifold is also indicated, which is the main cause of polariton group velocity renormalization. }
    \label{fig1}
\end{figure}
As such, $k_{\parallel}$ has discrete (but quasi-continuous) values of  $k_{\parallel} = \frac{2 \pi}{NL} k$, 
where the mode index is $k\in[-\frac{\mathcal{M}-1}{2},...0,...\frac{\mathcal{M}-1}{2}]$.
Diagonalizing $\hat{H}_\text{S}$ in the singly excited subspace leads to $2\mathcal{M}$ polariton states $|\pm_k\rangle$, with eigen-energies
\begin{align} \label{eq:wk_pm}
    \epsilon_{\pm k}=\hbar \omega_{\pm k} = \frac{\hbar}{2}(\omega_{k} + \omega_0) \pm \frac{\hbar}{2} \sqrt{(\omega_{k} - \omega_0)^2 + 4N g^2_{k}},
\end{align}
where $+$ and $-$ denotes the upper polariton (UP) and lower polariton (LP) branches, respectively. In addition, there are $N-\mathcal{M}$ dark states $|{\mathcal D}_k\rangle$ with energies $\hbar \omega_{{\mathcal D} k} = \hbar \omega_0$, which do not mix with photonic states and they form the dark polariton branch. 

Under the polariton representation, the system Hamiltonian in Eq.~\ref{eq:Hs_site} is expressed as $\hat{H}_\text{S} = \sum_{\mu,k} \hbar \omega_{\mu k} \hat{P}^\dagger_{\mu, k} \hat{P}_{\mu, k}$, where $\hat{P}^\dagger_{\mu, k}$, $\hat{P}_{\mu, k}$ are the polariton creation and annihilation operators for polariton state $k$ on polariton band $\mu$, respectively, and the band label $\mu \in \{+,-,\mathcal{D}\}$. Specifically, 
\begin{subequations}
\begin{align}
&\hat{P}^\dagger_{+, k} = \cos\Theta_{{k}} \hat{B}^\dagger_k + \sin\Theta_{{k}} \hat{a}^\dagger_k\\
&\hat{P}^\dagger_{-, k} = - \sin\Theta_{{k}} \hat{B}^\dagger_k + \cos\Theta_{{k}} \hat{a}^\dagger_k,
\end{align}    
\end{subequations}
where $\hat{B}^\dagger_k = (1 / \sqrt{N}) \sum_{n=1}^N e^{- i k_{\parallel}\cdot x_n} \hat{\sigma}^{\dagger}_n$ creates the collective bright excitons, and
\begin{equation}
\Theta_{{k}} = \frac{1}{2}\arctan \left( \frac{2 \sqrt{N} g_{k}}{\omega_{k} - \omega_0} \right) \in [0, \frac{\pi}{2})
\end{equation}
is the mixing angle. 
Details on the derivation in the polariton representation as well as the expressions of the polariton operators are provided in Supplementary Note 1. 
We present a schematic illustration of the model system above, as well as the polariton band structure, in Fig.~\ref{fig1}.
Without coupling to phonons, the polariton exhibits band-like transport characterized by the group velocity
\begin{align} \label{eq:vg_bare}
    v_{g, \pm}(k_\parallel) = d \omega_{\pm k} / dk_\parallel,
\end{align}
where the $k_\parallel$-dependence of $\omega_{\pm k}$ is carried by $\omega_k$ via Eq.~\ref{eq:wk-cavity}. 

The bath Hamiltonian $\hat{h}_\mathrm{B}$ describes the nuclear DOF, which we assume is a phonon environment that consists of a set of non-interacting harmonic oscillators, $\hat{h}_\text{B} = \sum_{n=1}^N \sum_{\alpha} \hbar \omega_\alpha \hat{b}^\dagger_{\alpha, {n}} \hat{b}_{\alpha, {n}}$, where $\hat{b}_{\alpha, n}$, $\hat{b}^\dagger_{\alpha, n}$ are the $\alpha_\mathrm{th}$ bosonic bath phonon annihilation and creation operators in the $n_\mathrm{th}$ molecule with phonon frequency $\omega_\alpha$. Furthermore, $\hat{H}_\mathrm{SB}$ describes the exciton-phonon interaction $\hat{H}_\text{SB} = \sum_{n=1}^N \hat{\sigma}_n^{\dagger} \hat{\sigma}_n \otimes \sum_{\alpha} c_\alpha (\hat{b}_{\alpha, {n}} + \hat{b}^\dagger_{\alpha, {n}})$, where $c_\alpha$ is the exciton-phonon coupling strength. We assume the coupling strength is identical for all excitons and $c_\alpha$ is therefore independent of the label $n$. 
Based on the Caldeira-Leggett model~\cite{Caldeira_Leggett_1983, Nitzan}, the baths as well as their interactions with the system are described by the spectral density 
\begin{equation}
J(\omega) = \frac{\pi}{\hbar} \sum_{\alpha} c^2_{\alpha} \delta(\omega - \omega_\alpha),
\end{equation}
and $\lambda = (1/\pi) \int_0^{+\infty} d\omega~J(\omega) / \omega  = \sum_\alpha c^2_\alpha / \omega_\alpha$ is the reorganization energy. 

We further introduce the Fourier transform of the bath phonon operators $\hat{b}_{\alpha, {k}} = (1 / \sqrt{N}) \sum_{n = 1}^N e^{i k_\parallel \cdot x_n} \hat{b}_{\alpha, {n}}$.
Using these transforms, the bath Hamiltonian is expressed as $\hat{h}_\text{B} = \sum_k \sum_\alpha \hbar \omega_\alpha \hat{b}^\dagger_{\alpha, {k}} \hat{b}_{\alpha, {k}}$, and the polariton-phonon interaction Hamiltonian is given by
\begin{equation}
\hat{H}_\text{SB}= \sum_{\mu,k,\nu,k'} \zeta_{\mu k} \cdot \zeta_{\nu k'} \hat{P}^\dagger_{\mu, k} \hat{P}_{\nu, k'} \sum_\alpha \frac{c_\alpha}{\sqrt{N}} (\hat{b}_{\alpha, {k - k'}} + \hat{b}^\dagger_{\alpha, {k'-k}}), 
\end{equation}
where the band labels $\mu, \nu \in \{+, -, \mathcal{D}\}$, and $\zeta_{\mu k}$ is a state-dependent coefficient that characterizes the matter fraction of the polariton state, with $\zeta_{+k}=\cos\Theta_{{k}}$ and $\zeta_{-k}=\sin\Theta_{{k}}$.
The $\zeta_{+k}$ and $\zeta_{-k}$ are commonly referred to as the Hopfield coefficients \cite{Deng_RMP2010, chng2024mechanism, qiu2021molecular}, and we note that $\zeta_{\mathcal{D}k}=1$. These polariton-phonon interactions will modify the polariton band structure, and will, in turn, affect the polariton transport properties such as the group velocity in Eq.~\ref{eq:vg_bare}. 

{\bf Theory}. We derive the expression for $v_{g}$-renormalization using the {\it equilibrium} Green's functions at finite temperature. We restrict our discussions on polariton transport in the weak exciton-phonon coupling regime and the band-like transport regime~\cite{Milan_NC2023, Tal_NM2023}. The single-particle Green's function of the polaritons at finite temperature is expressed as follows~\cite{Mahan},
\begin{align} \label{eq:GF-polariton}
    G_{\mu, k}(t) \equiv -i \theta(t) \langle \hat{P}_{\mu, k}(t) \hat{P}^\dagger_{\mu, k}(0) \rangle,
\end{align}
where $\theta(t)$ is the Heaviside step function, the time-dependence of the operators read as $\hat{P}_{\mu, k}(t) = e^{\frac{i}{\hbar} \hat{H} t} \hat{P}_{\mu, k}(0) e^{- \frac{i}{\hbar} \hat{H} t}$, and $\langle \hat{A} \rangle \equiv \text{Tr}[\hat{A} e^{-\beta \hat{H}}] / \text{Tr}[e^{-\beta \hat{H}}]$ denotes the thermal average under finite temperature $\beta \equiv 1 / (k_\text{B} T)$, where $k_\text{B}$ is the Boltzmann constant. Similarly, one defines the Green's function of the phonons as $D_q(t) \equiv - i \sum_\alpha (c^2_\alpha / N) \cdot \langle \theta(t) \hat{b}_{\alpha, q} (t) \hat{b}^\dagger_{\alpha, q} (0) + \theta(-t) \hat{b}^\dagger_{\alpha, -q} (t) \hat{b}_{\alpha, -q} (0) \rangle$. The Green's function in Eq.~\ref{eq:GF-polariton} can be determined by the self-consistent Dyson equation in the time domain as~\cite{Mahan}
\begin{align} \label{eq:Dyson_eq_t}
    \Big(i\hbar \frac{\partial}{\partial t} - \epsilon_{\mu k}\Big) G_{\mu, k}(t) - \int_0^t d\tau~ \Sigma_{\mu, k}(t-\tau) G_{\mu, k}(\tau) = \delta(t),
\end{align}
where $\Sigma_{\mu, k}(t)$ is the self-energy, and $\epsilon_{\mu k} = \hbar \omega_{\mu k}$ is the bare polariton energy. Eq.~\ref{eq:Dyson_eq_t} is recast in the frequency domain as
\begin{align} \label{eq:Dyson_eq_w}
    \mathcal{G}^{-1}_{\mu, k}(\omega) = \hbar (\omega - \omega_{\mu k} + i\eta ) - \Sigma_{\mu, k}(\omega),
\end{align}
where $\mathcal{G}_{\mu, k}(\omega)$ is the Fourier transform of $G_{\mu, k}(t)$, and we take $\eta \to 0_+$. To obtain the polariton band renormalization, we further define the renormalized polariton energies $\tilde{E}_{\mu k} = E_{\mu k} + i\Gamma_{\mu k}$ and plug it into Eq.~\ref{eq:Dyson_eq_w}, arriving at the expression~\cite{Giustino_2017}
\begin{subequations} \label{eq:energy_sc}
\begin{align}
    E_{\mu k} &= \hbar \omega_{\mu k} + \text{Re}[\Sigma_{\mu, k}(\tilde{E}_{\mu k} / \hbar)], \\
    \Gamma_{\mu k} &= \text{Im}[\Sigma_{\mu k}(\tilde{E}_{\mu k} / \hbar)],
\end{align}
\end{subequations}
which has to be solved self-consistently for $E_{\mu k}$ and $\Gamma_{\mu k}$.
Consequently, $E_{\mu k}$ is the renormalized polariton band, and the renormalized polariton group velocity is obtained via $\tilde{v}_{g, \pm}(k_\parallel) = (1 / \hbar) d E_{\pm k} / d k_\parallel$, which leads to
\begin{align} \label{eq:vg_renorm}
    \tilde{v}_{g, \pm}(k_\parallel) = v_{g, \pm}(k_\parallel) + \frac{1}{\hbar} \frac{d}{d k_\parallel} \text{Re}[\Sigma_{\pm, k}(\tilde{E}_{\pm k} / \hbar)].
\end{align}
The second term in the right-hand side of Eq.~\ref{eq:vg_renorm} characterizes the modification of the polariton group velocity due to polariton-phonon interaction. We hypothesize that this term is the main cause of the renormalization of $v_{g}$~\cite{Milan_NC2023,chng2024transport}.

In most cases, Eq.~\ref{eq:energy_sc} cannot be solved exactly and approximations are needed to obtain the self-energy in a closed form. Here, we derive the leading contribution to polariton band renormalization using the standard tools of diagrammatic perturbation theory. The first-order self-energy is expressed as~\cite{Mahan, Reichman_2022_1, Reichman_2022_2} 
\begin{align} \label{eq:SE-2nd}
    \Sigma^{(1)}_{\mu, k}(t) = i \zeta^2_{\mu k} \sum_{\nu, k'} \zeta^2_{\nu k'} \cdot D^{(0)}_{k - k'}(t) G^{(0)}_{\nu, k'}(t),
\end{align}
where $G^{(0)}_{\pm, k}(t) = -i \theta(t) e^{-i \omega_{\pm k} t}$ and $G^{(0)}_{\mathcal{D}, k}(t) = -i \theta(t) e^{-i \omega_0 t}$ are the non-interacting Green functions of the polaritons, and the low-temperature limit is taken because $\epsilon_{\mu k} \gg k_\text{B}T$. Further, $D^{(0)}_{k - k'}(t)$ is the free phonon propagator under finite temperature, and is expressed as
\begin{equation}
    D^{(0)}_{q}(t) = -i \sum_\alpha \frac{2 c^2_\alpha}{N} [(1 + \overline{n}_\alpha) e^{-i\omega_\alpha |t|} + \overline{n}_\alpha e^{i\omega_\alpha |t|}],
\end{equation}
where $D^{(0)}_{q}(t)$ is independent of $q$ (see right-hand side of Eq.~\ref{eq:SE-2nd}), $\overline{n}_\alpha = 1 / (e^{\beta \hbar \omega_\alpha} - 1)$ is the Bose-Einstein distribution function, and the factor of 2 results from the degeneracy of the bath modes such that $\omega_{\alpha, q} = \omega_{\alpha, - q} = \omega_{\alpha}$. Eq.~\ref{eq:SE-2nd} is the Fan-Migdal self-energy~\cite{Giustino_2017}, and when substituted in Eq.~\ref{eq:vg_renorm} leads to the following expression for the modified polariton bands
\begin{equation} \label{eq:wk_modified_2}
E^{(2)}_{\mu k} = \hbar \omega_{\mu k} + \zeta^2_{\mu k} \cdot \sum_{\nu, k'} \sum_\alpha \zeta^2_{\nu k'} \cdot \frac{2 c^2_\alpha}{N} \cdot  \Xi_{\mu k,\nu k'}(\omega_{\alpha}),
\end{equation}
where $\Xi_{\mu k,\nu k'}(\omega_{\alpha})$ is the real part of the polarizability and is given by
\begin{align}\label{Xi}
&\Xi_{\mu k,\nu k'}(\omega_{\alpha})\\
&=\text{Re} \Big[\frac{1 + \overline{n}_\alpha}{\omega_{\mu k} - \omega_{\nu k'} - \omega_\alpha + i\eta} + \frac{\overline{n}_\alpha}{\omega_{\mu k} - \omega_{\nu k'} + \omega_\alpha + i\eta} \Big].  \notag
\end{align}
A detailed derivation of Eq.~\ref{Xi} is provided in Supplementary Note 2.
For continuous spectral density functions, the summation over the phonon modes $\alpha$ in Eq.~\ref{eq:wk_modified_2} can be written as an integral in terms of $J(\omega)$ (see Supplementary Note 3). We note that the band modification can also be obtained directly from the total Hamiltonian using Rayleigh-Schr\"odinger perturbation theory~\cite{Giustino_2017}, by treating $\hat{H}_\text{SB}$ as perturbative interactions that cause 2nd order energy corrections (that scatter $|-,k\rangle$ to dark states then scatter back). This derivation is provided in Supplementary Note 2D, with the results identical to Eq.~\ref{eq:wk_modified_2} (with $\eta=0$).

In this work, we focus on the LP's $v_{g}$ renormalization, which is dominated by scattering to the dark exciton states (a total of $N-{\mathcal M}$ of them), as opposed to scattering to the $\mathcal{M}$ LP and $\mathcal{M}$ UP states, because $N - {\mathcal M} \gg 2{\mathcal M}$. Thus, one can explicitly perform the summation over $k'$ that only includes the dark exciton contributions, with $\sum_{k'}f(\omega_{\nu k'})\approx (N-\mathcal{M}) f (\omega_{0})$, and the $N-\mathcal{M}$ factor will cancel with $1/N$ in Eq.~\ref{Xi} under the large $N$ limit. This cancellation also indicates that in simulations, as long as one can keep $N-\mathcal{M}/N\to 1$, one should expect the same converged results, and the detailed choice of $N$ or $\mathcal{M}$ does not matter that much (assuming sufficient resolution of the polariton wavepacket in the spatial and $k$-space).

With the above considerations, the renormalized LP group velocity becomes
\begin{equation}\label{vgLP}
\tilde{v}_{g,-}=v_{g,-}+ \frac{d}{d k_\parallel} \left[|C_{k}|^2\sum_{\alpha} 2c^2_{\alpha} \cdot \Xi_{-k,0}(\omega_{\alpha})\right],
\end{equation}
where the Hopfield coefficient $|C_{k}|^2$ is expressed as
\begin{equation}
|C_{k}|^2 = \sin^2 \Theta_k = \frac{1}{2} \Big[1 + \frac{\omega_{k}-\omega_{0}}{\sqrt{(\omega_{k}-\omega_{0})^2 + 4Ng^2_{k}}} \Big],\nonumber
\end{equation}
which characterizes the matter fraction of the LP.
Further, $\Xi_{- k, 0}(\omega_{\alpha})$ only considers the dark exciton contribution, and is expressed as
\begin{align}\label{XiLP}
&\Xi_{- k, 0}(\omega_{\alpha}) =\frac{\overline{n}_\alpha \cdot (\omega_{\alpha}-\Delta \omega_{-k})}{(\omega_{\alpha}-\Delta \omega_{-k})^2+ \eta^2} - \frac{1 + \overline{n}_\alpha}{\omega_{\alpha}+\Delta \omega_{-k}}, 
\end{align}
where $\Delta\omega_{-k}=\omega_{0}-\omega_{- k}> 0$ is the energy gap between the dark exciton states and the LP band at $k_{\parallel}=\frac{2 \pi}{NL} k$. 
Eq.~\ref{vgLP} provides an analytic expression of the LP group velocity based on the current theory. It predicts that the magnitude of the $v_{g}$ renormalization will depend linearly on $\lambda$ [through $c^2_\alpha$], and also predicts that $v_{g}$ is sensitive to $C_{k}$ and temperature [through $\overline{n}_\alpha$]. 
Further taking the $\eta\to0$ limit of Eq.~\ref{XiLP}, one can analytically express Eq.~\ref{vgLP} as
\begin{align}\label{vg-final}
\Delta v_{g,-}&\equiv\tilde{v}_{g,-}-v_{g,-}\\
&=-\frac{d}{d k_{\parallel}} \Big[|C_{k}|^2 \sum_{\alpha}2 c_{\alpha}^2 \omega_{\alpha}\frac{\Delta \omega_{-k}\cdot (2 \overline{n}_\alpha + 1) -\omega_{\alpha}}{\Delta\omega^2_{-k}-\omega^2_{\alpha}}\Big]. \nonumber
\end{align}
In most experiments, the LP initial excitation is in a region $\Delta\omega_{-k}\gg \omega_{\alpha}$, thus $\Xi_{- k, 0}(\omega_{\alpha})$ is negative. For a broad range of phonon frequencies, the high-frequency phonon makes a positive contribution to $\Xi_{- k, 0}(\omega_{\alpha})$, but the overall results should still be dominated by the low-frequency phonons, making $\Xi_{- k, 0}(\omega_{\alpha})$ negative. Note that Eq.~\ref{vgLP} is only valid when dark excitons dominate the sum in Eq.~\ref{eq:wk_modified_2}. Nevertheless, one is able to derive simpler analytic answers from Eq.~\ref{eq:wk_modified_2} or Eq.~\ref{vgLP} under different regimes of spectral densities $J(\omega)$ or temperatures. 

{\bf Mechanistic Picture}. We want to comment on the mechanistic picture suggested by Eqs.~\ref{vg-final} and ~\ref{eq:wk_modified_2}. The LP group velocity renormalization occurs mainly due to the presence of the dark states as a virtual scattering state. The transition from LP to all dark states, and scattering back to the LP  ($|-,k\rangle \to |\mathcal{D}\rangle\to |-,k\rangle$) leads to the reduction of the group velocity, which can be understood as the perturbative energy correction up to second order. Indeed, the overall scaling of $\Delta v_{g,-}\propto 1/\Delta \omega_{-k}$. This scaling means that even with large light matter detunings, such that the dark states are never appreciably populated from the LP, these dark states still act like virtual states, such that their perturbative presence will lead to energy correction of LP and hence $v_{g}$ renormalization. In this sense, we can classify the physical picture predicted by Eq.~\ref{vg-final} as the super-exchange mechanism, where the dark exciton states act like virtual states to the super-exchange population with LP. For small light-matter detuning (such as in Ref.~\citenum{Musser_2022}), the LP might be able to transfer the population to the dark states. For large light-matter detuning, dark states will only be virtually populated and thus will not be detected spectroscopically, as experimentally observed under resonant excitation of the LP in Ref.~\citenum{Milan_NC2023}. We also note that the mechanism is also akin to the Raman scattering process, which is evidenced by the expression of $\Xi_{\mu k,\nu k'}(\omega_{\alpha})$ in Eq.~\ref{Xi}. In fact, Eq.~\ref{eq:wk_modified_2} is the Raman-type polarizability in the frequency domain, which is the well-known Kramers-Heisenberg-Dirac (KHD) expression~\cite{Kramers1925, Dirac1927, Dirac1927-paper2, Tannor-book}, but now with temperature dependence (because the interaction is $\hat{H}_\mathrm{SB}$, which is temperature dependent, and not the dipole interaction with the field in the original KHD expression). Supplementary Note 2D clearly shows how the $\hat{H}_\mathrm{SB}$ term mediates the transition from LP to dark states and back to LP bands. As such, the $v_{g}$-renormalization can also be described as a phonon-mediated Raman-type scattering process, which is a non-resonant process. A schematic illustration is provided in Fig.~\ref{fig1}b. 
\begin{figure*}
    \centering
    \includegraphics[width=0.7\linewidth]{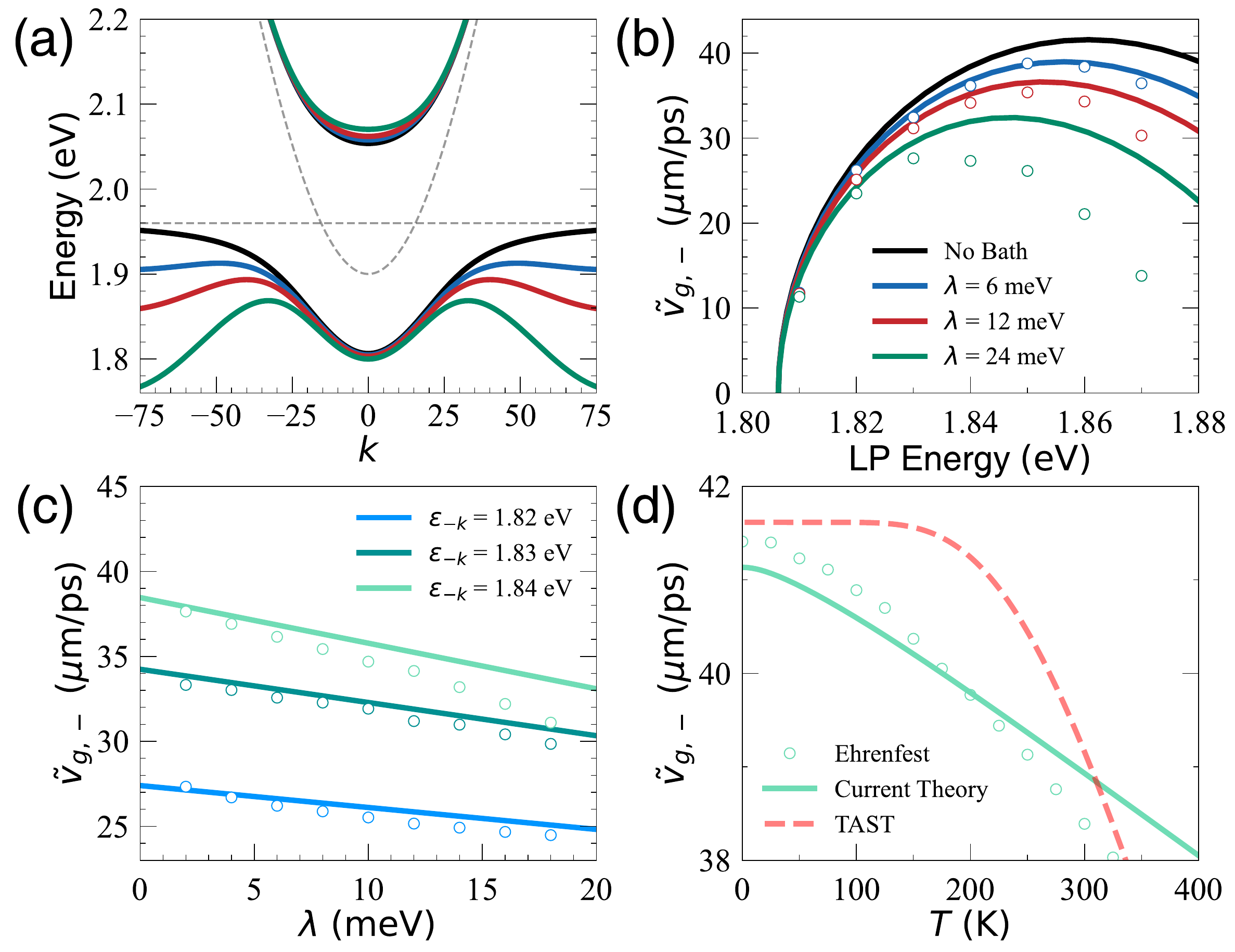}
    \caption{{\bf Polariton band structure modification and group velocity renormalization due to polariton-phonon interaction}. ({\bf a}) Modified polariton band structure under different $\lambda$. ({\bf b}) Group velocity of the LP branch $\tilde{v}_{g, -}$ under different $\lambda$. ({\bf c}) Scaling relation of the LP group velocity $\tilde{v}_{g, -}$ with $\lambda$. ({\bf d}) Temperature-dependence of the LP group velocity $\tilde{v}_{g, -}$ at LP energy $\epsilon_{-k} =$ 1.86 eV and $\lambda = $ 6 meV. Theoretical results using Eq.~\ref{eq:wk_modified_2} (solid lines) are compared to Ehrenfest dynamics simulations (open circles). }
    \label{fig2}
\end{figure*}

{\bf Numerical Results}. To quantitatively examine the accuracy of the above theory (Eq.~\ref{eq:wk_modified_2}, or the corresponding $\tilde{v}_{g,-}$), we perform quantum dynamics simulations for the GHTC model Hamiltonian using the Ehrenfest method~\cite{chng2024transport}, and verify various scaling relations and predictions made by the theory. For the system Hamiltonian, we chose the exciton energy $\hbar \omega_0 = 1.96$ eV, the cavity frequency $\hbar \omega_\text{c} = 1.90$ eV, and the collective light-matter coupling strength $\sqrt{N} g_\text{c} = 120$ meV. Details of the models and computations are provided in Supplementary Note 4, with a brief summary provided in the Methods section.

Fig.~\ref{fig2}{\bf a} presents the modified polariton band structure with different $\lambda$. One observes that the modification of $v_g$ increases as $\lambda$ and the matter fraction increases. For the LP branch, the second term in Eq.~\ref{eq:wk_modified_2} is negative, which effectively provides an {\it attractive} interaction between polaritons (mediated by phonons) and decreases the LP energy. Similarly, the energy increases for the UP branch. Since $\zeta^2_{\mu k}$ is the matter fraction of the polariton branch, it is straightforward to see that as $k_\parallel$ increases, $\zeta^2_{- k}$ increases with larger matter fraction, thus providing more modifications to the LP band. 
The modified polariton band structure consequently leads to polariton group velocity renormalization. 

Fig.~\ref{fig2}{\bf b} presents the LP group velocity at different energies (see Fig.~\ref{fig2}{\bf a}) and for different $\lambda$, where the theoretical results using Eq.~\ref{eq:wk_modified_2} are compared to quantum dynamics simulations (open circles). One sees that as $\lambda$ increases, the magnitude of the group velocity renormalization increases (from the blue curve to the green curve), further deviating from the derivative of the LP band, $v_{g}$ (black solid curve). Further, as the LP energy increases, the matter character of the LP state $|C_{k}^2|$ also increases, which further reduces the group velocity. For all cases, the theory agrees very well with the numerical simulations for small $\lambda$ ($<$ 12 meV). However, for larger $\lambda$, the polariton-phonon interaction enters the non-perturbative regime, and the first-order self-energy level theory in Eq.~\ref{eq:wk_modified_2} becomes inadequate. As a result, the theory gradually deviates from numerical simulations, as expected. Nevertheless, the theory describes the overall semi-quantitative trend of the data from the simulation. 

Fig.~\ref{fig2}{\bf c} presents the scaling relation of the LP group velocity $\tilde{v}_{g, -}$ (c.f. Eq.~\ref{vgLP}) as a function of $\lambda$, which characterizes the modification to the LP group velocity by the polariton-phonon interaction. Importantly, the theory in Eq.~\ref{vgLP} predicts that this renormalization magnitude is proportional to $c^2_{\alpha}$ and thus $|\Delta v_{g,-}| = |\tilde{v}_{g, -} - v_{g, -}| \propto \lambda$. Fig.~\ref{fig2}c presents $\tilde{v}_{g, -}$ versus $\lambda$ at different LP energies. We observe that $\tilde{v}_{g, -}$ scales linearly with $\lambda$, and the slope increases as the matter fraction increases. It is clear from Eq.~\ref{eq:wk_modified_2} that the polariton band structure (or group velocity) modification is proportional to $\lambda$ due to its quadratic dependence on $c_\alpha$. 
The results obtained from quantum dynamics simulations agree quite well with the theory, especially for cases with small $\lambda$ and matter fractions. As $\lambda$ and matter fraction increase, the Ehrenfest results gradually deviate from the theory and show a nonlinear dependence on $\lambda$, due to non-perturbative effects; see the $\epsilon_{-k} = $ 1.84 eV (shallow green) curve for example. Nevertheless, the semi-quantitative trend is always captured by the theory, and we stress that there are {\it no free parameters} in the current theory. Furthermore, our quantum dynamics simulation is based on the Ehrenfest MQC approximation which may lead to inaccurate results when $\lambda$ is large. Future efforts are needed to evaluate $v_{g}$ in the large $\lambda$ regime using more accurate quantum dynamics approaches. 

Fig.~\ref{fig2}{\bf d} presents the temperature dependence of the polariton group velocity renormalization. Fig.~\ref{fig2}{\bf d} presents $\tilde{v}_{g, -}$ versus $T$ at LP energy $\epsilon_{-, k} =$ 1.86 eV and $\lambda =$ 6 meV. 
From a theoretical standpoint, the temperature dependence is mainly carried by the Bose-Einstein distribution function in Eq.~\ref{eq:wk_modified_2} which is nonlinear in $T$. In particular, under the high-temperature limit ($\hbar \omega_{\alpha} \ll k_\text{B} T$ for all $\omega_{\alpha}$), the Bose-Einstein distribution function can be approximated as $\overline{n}_\alpha \approx k_\text{B} T / (\hbar\omega_\alpha)\propto T$. As a result, the modification of the polariton band structure (or group velocity) is proportional to $T$. At temperatures near 300 K, the parameters we used satisfy the high-temperature limit, thus $\Delta v_{g, -}$ scales linearly with $T$. In the Ehrenfest dynamics simulations, the nuclear quantum effect comes from the Wigner distribution of the nuclear thermal density only, which does not give accurate results in the very low-temperature regime. Considering this, we place greater confidence in the analytic theory, which should be accurate under the low $\lambda$ and $T\to 0$ limits because of the perturbative treatment. Nevertheless, both the current theory (solid green line) and the numerical simulation agree reasonably well across all temperature regimes. Overall, the theory and simulations predict that $v_{g,-}$ decreases as $T$ increases. This is because when $T$ increases, the phonon fluctuations cause transitions from the LP band to the dark exciton states, thus reducing the group velocity. We also want to emphasize that there is {\it no free parameter} in current theory to predict the temperature dependence.

Note that a phenomenological expression has previously been proposed based on the thermally activated scattering theory (TAST)~\cite{Tal_NM2023}, due to scattering from $|-_k\rangle$ to the dark states, resulting in the following expression for the group velocity renormalization 
\begin{equation}\label{TAST}
\tilde{v}_{g,-}=\frac{v_{g,-}}{1+G \cdot e^{-\beta\hbar\Delta\omega_{-k}}},
\end{equation}
where $G$ is a free parameter. See Supplementary Note 5, as well as Supplementary Information S3 in Ref.~\citenum{Tal_NM2023} for further details. The TAST is based on the idea that transport depends on the proportion of time the system spends in the LP band relative to the dark states, resulting in a temperature-dependent modification of $v_g$ that is sensitive to the energy gap $\Delta\omega_{-k}$. Although the TAST makes intuitive sense (and aligns with findings from our microscopic theory), we found that Eq.~\ref{TAST} does not give the correct temperature dependence when $G$ is treated as a temperature-independent parameter. In Fig.~\ref{fig2}, the result from TAST is plotted as the red dashed curve, with a fitting parameter $G=3.0$ to reproduce the correct value of $\tilde{v}_{g,-}$ at $T=300$ K. One sees that it does not give the correct $T$-dependence across a broad range of temperatures unless one further chooses a $T$-dependent $G$ parameter. The reason TAST fails to reproduce an accurate $T$-dependence is because the expression from TAST scales as $1/(1+e^{-\beta\hbar\Delta \omega_{-k}})$, whereas the microscopic theory 
in Eq.~\ref{vgLP} posits that the temperature dependence is $\overline{n}_\alpha\approx e^{-\beta\hbar\omega_{\alpha}}$ under the low-temperature limit when $\hbar \omega_{\alpha}\gg k_\mathrm{B}T$, and $\overline{n}_\alpha \approx k_\text{B} T / (\hbar\omega_\alpha)$ under the high-temperature limit when $\hbar \omega_{\alpha}\ll k_\mathrm{B}T$. Additionally, TAST assumes that the transition between the LP band and dark exciton states follows Boltzmann statistics, whereas, in our current theory, the phonons obey the Bose-Einstein statistics, which mediate the (virtual) transitions between the LP band and the dark states. Our microscopic theory also predicts that $\Delta v_{g,-}$ should depend on $\Delta \omega_{-k}$, but this dependence (in Eq.~\ref{vgLP}) is not in the Boltzmann factor.



\section*{Conclusions}
We developed a microscopic theory that successfully explains the renormalization of polariton group velocity due to polariton-phonon interactions. We analyze a theoretical model based on the GHTC Hamiltonian, which comprises of $N$ identical copies of molecular systems consisting of excitons and phonons that are collectively coupled to $\mathcal{M}$ cavity modes which satisfy some dispersion relation. The theory uses a diagrammatic perturbative treatment to the {\it equilibrium} Green's function of the polaritons, revealing how exciton-phonon interactions renormalize the LP band and thus reduce the group velocity in polariton transport. Crucially, the theory captures the $\lambda$ and $T$ dependence of the $v_{g}$ renormalization magnitude and semi-quantitatively agrees with results from quantum dynamics simulations. We emphasize that there is no free parameter in our microscopic theory, and every quantity is derived from the microscopic light-matter interaction Hamiltonian. 

We expect the theory will eventually break down with increasing $\lambda$ and matter fraction such that the system enters into the non-perturbative regime. However, for $\lambda\le k_\mathrm{B}T$, the analytic theory almost quantitatively agrees with the numerical results. Although the theory does not capture transient non-equilibrium dynamical behaviors in the short-time regime, it yields quantitatively accurate answers compared to numerical simulations that do include all transient non-equilibrium effects. This strongly suggests that the LP $v_{g}$ renormalization is largely dictated by the renormalization of the LP band due to phonons and is less sensitive to the transient, non-equilibrium dynamics. 


Our theory yields several predictions regarding the scaling relation with matter fraction $|C_{k}|^2$, phonon bath reorganization energy $\lambda$, temperature {\it, etc}, and these have been verified through our quantum dynamics simulations. These predictions can, in principle, be verified with experiments~\cite{Milan_NC2023, Tal_NM2023, Musser_2022}. The theory is simple enough to be extended to multidimensional systems with multiple dispersive matter bands and phonons, such as semiconductor materials. It is also feasible to implement our theory along with {\it ab initio} simulations~\cite{Giustino_2017}.


\section*{Methods}
{\bf Numerical Evaluation of Eq.~\ref{eq:wk_modified_2}}. We assume a Drude-Lorentz form for the phonon bath spectral density $J(\omega) = \frac{\pi}{\hbar} \sum_{\alpha} c^2_{\alpha} \delta(\omega - \omega_\alpha) = 2\lambda \omega_f \omega / (\omega^2 + \omega^2_f)$, where $\lambda$ is the reorganization energy, and $\omega_f$ is the bath characteristic frequency. We adopt an efficient and commonly used type of bath discretization procedure~\cite{Makri_2017}, which discretizes the spectral density with equal intervals in $\lambda$ (instead of in frequency), with 
\begin{equation*}
    \omega_\alpha = \omega_f \tan [\frac{\pi}{2} (1 - \frac{\alpha}{N_b + 1})],~~~~~~
    c_\alpha = \sqrt{\frac{\lambda \omega_\alpha}{N_b + 1}},
\end{equation*}
where $\alpha = 1, \cdots, N_b$, and $N_b$ is the number of bath modes. Here $N_{b}=10^4$ is used to evaluate Eq.~\ref{eq:wk_modified_2} to generate converged results, and the infinitesimal imaginary term in Eq.~\ref{Xi} is taken as $\eta =$ 1 meV. 
The value of $\tilde{v}_{-,g}$ is directly obtained by numerically differentiating $E^{(2)}_{\mu k}$ in Eq.~\ref{eq:wk_modified_2}. Note that one can adopt a smaller $\eta$ value in numerical calculations, but it then requires an even larger $N_b$ to reach to convergence. 

{\bf Quantum Dynamics Simulations}. We use the mean-field Ehrenfest dynamics~\cite{hu2024trajectory}  to propagate the quantum dynamics of polariton transport. The transport dynamics mainly occur in the single excitation subspace, defined as follows
\begin{subequations}\label{basis}
\begin{align}
\ket{{E}_n} &=  \ket{{e}_n}\bigotimes\limits_{m\neq n}\ket{{g}_m}\bigotimes_{k} |0_{{k}}\rangle\nonumber\\
\ket{{k}} &= |G\rangle \bigotimes\limits_{{k'} \neq {k}} |0_{{k'}}\rangle \otimes \ket{1_{k}},\nonumber
\end{align}
\end{subequations}
where $\ket{{E}_n}$ is the singly excited state for the $n_\mathrm{th}$ molecule located at $x_{n}$, $\ket{k}$ is the one-photon-dressed ground state with wave-vector ${k_\parallel}= \frac{2 \pi}{NL} k$, and $|G\rangle=\bigotimes\limits_{n}\ket{{g}_n}\bigotimes\limits_{\alpha}\ket{0_{k}}$ represents the common ground state for the hybrid system. 
We describe the time-dependent quantum state in the exciton-photon subspace as
\begin{align}\label{eq:polaritonwavepacket}
|\psi(t)\rangle &= \sum_{n}^{N} c_{n}(t)\ket{E_n} +  \sum_k c_k (t) \ket{{k}},\nonumber
\end{align}
where $c_n (t)$ and $c_k (t)$ are the excitonic and photonic expansion coefficients respectively. The polariton quantum dynamics for $|\psi(t)\rangle$ is propagated by solving the time-dependent Schr\"odinger equation (TDSE),
\begin{equation}\label{TDSE}
i\hbar \frac{\partial}{\partial t}|\psi (t)\rangle=\hat{H}_\mathrm{Q} ({\bf R})|\psi (t)\rangle, \nonumber
\end{equation}
where $\hat{H}_\mathrm{Q}=\hat{H}_\mathrm{S}+\hat{H}_\mathrm{SB}({\bf R})$. The bath nuclear DOF ${\bf R}$, on the other hand, is propagated classically using Hamiltonian's equations of motion (EOM), governed by the time-dependent mean-field force 
\begin{equation}\label{nuclearForce}
    {\bf F} =-\nabla_{\bf R} \left[\langle \psi(t)|\hat{H}_\mathrm{SB}({\bf R})|\psi(t)\rangle+{h}_\mathrm{B}({\bf R})\right]. \nonumber
\end{equation}
The polariton group velocity $\tilde{v}_{g}$ is computed by tracking the wavefront of the LP polariton wavepacket using the same method reported in previous works~\cite{Milan_NC2023, chng2024transport}, with details provided in Supplementary Note 4.

\textbf{Simulation Details.} For all quantum dynamics simulations, we use $N = 10^4$ molecules and $\mathcal{M} = 10^2$ cavity modes, keeping the ratio of $N/\mathcal{M}\approx 35$. More details about the precise number of molecules and modes for each parameter regime explored in Fig.~\ref{fig2} are provided in Supplementary Note 4. A total of $N_{b}=35$ phonon modes were sampled based on the same equal-$\lambda$ procedure mentioned above. The total light-matter coupling strength is set to $\sqrt{N}g_\mathrm{c} = 120$ meV. All results are obtained with an ensemble of 500 independent trajectories. Convergence tests are performed with up to 1000 trajectories. The nuclear time step is $\Delta t=2.5$ fs, where during each nuclear propagation, there are 100 electronic propagation steps with a time step $dt=0.025$ fs. The nuclear EOM is numerically integrated with the velocity verlet algorithm and the TDSE is solved with the Runge-Kutta-4 algorithm. \\

\section*{Acknowledgements}
This work was supported by the National Science Foundation Award under Grant No. CHE-2244683, the Air Force Office of Scientific Research under AFOSR Award No. FA9550-23-1-0438, as well as by the University of Rochester PumpPrimer II funding. M.D. acknowledges support from the Office of Naval Research under Grant No. N00014-25-1-2079.  W.Y. appreciates the support of the Moses Passer Graduate Fellowship at the University of Rochester. P.H. appreciates the support of the Cottrell Scholar Award (a program by the Research Corporation for Science Advancement). Computing resources were provided by the Center for Integrated Research Computing (CIRC) at the University of Rochester. 

\section*{Data Availability}
\noindent  The data that support the plots within this paper and other findings of this study are available from the corresponding authors upon a reasonable request.
\\

\section*{Code availability}
The source code that supports the findings of this study is available from the corresponding author upon reasonable request. 
\\

\section*{Author contributions}
\noindent All authors designed the research. W.Y. and P.H. conducted theoretical derivations. W.Y. performed numerical evaluations of the theoretical expressions, B.X.K.C. performed the quantum dynamics simulations of polariton transport. M.D. and P.H. developed the mechanistic picture of group velocity re-normalization. W.Y., B.X.K.C., M.D., and P.H. wrote the manuscript. 

\section*{Competing interests}
\noindent The authors declare no competing interests.

\section*{Additional Information}
\noindent {\bf Supplementary information} The online version contains supplementary material available at [url]. \\

\noindent {\bf Correspondence} and requests for materials should be addressed to Pengfei Huo.


\providecommand{\noopsort}[1]{}\providecommand{\singleletter}[1]{#1}%

\end{document}